\documentclass[hypertexnames=false,twocolumn,aps,prx,nobibnotes]{revtex4-2} 
\usepackage{array, makecell}

\usepackage{fancyref}
\usepackage{hyperref}
\usepackage{dcolumn}
\usepackage{bm}
\usepackage{tikz}
\usepackage{pgfplots}
\pgfplotsset{compat=1.18}
\tikzset{>=latex}
\usepackage{amsfonts}

\usepackage{booktabs}

\usepackage[mode=buildnew]{standalone}

\usepackage{lipsum}
\usepackage{graphicx} 
\usepackage{standalone}
\usepackage[caption=false]{subfig}
\usepackage[percent]{overpic}
\usepackage{verbatim}
\usepackage{tcolorbox}

\usepackage{physics} 
\usepackage{enumerate}
\usepackage{dsfont}
\usepackage{xcolor}

\usepackage[normalem]{ulem}
\usepackage[capitalise]{cleveref}

\usepackage{lipsum}

\begin{document}

\title{Quantum low-density parity-check codes for erasure-biased atomic quantum processors}

\author{Laura Pecorari}
\email{lpecorari@unistra.fr}
\affiliation{University of Strasbourg and CNRS, CESQ and ISIS (UMR 7006), aQCess, 67000 Strasbourg, France}

\author{Guido Pupillo}
\email{pupillo@unistra.fr}
\affiliation{University of Strasbourg and CNRS, CESQ and ISIS (UMR 7006), aQCess, 67000 Strasbourg, France}

\date{\today}

\begin{abstract} 
    Identifying the best families of quantum error correction (QEC) codes for near-term experiments is key to enabling fault-tolerant quantum computing. Ideally, such codes should have low overhead in qubit number, high physical error thresholds, and moderate requirements on qubit connectivity to simplify experiments, while allowing for high logical error suppression. Quantum Low-Density Parity-Check (LDPC) codes have been recently shown to provide a path towards QEC 
    with low qubit overhead and small logical error probabilities.  
    Here, we demonstrate that when the dominant errors are erasures -- as can be engineered in different quantum computing architectures -- quantum LDPC codes additionally provide high thresholds and even stronger logical error suppression in parameter regimes that are accessible to current experiments. Using large-scale QEC numerical simulations, we benchmark the performance of two families of high-rate quantum LDPC codes, namely Clifford-deformed La-cross codes and Bivariate Bicycle codes, under a noise model strongly biased towards erasure errors. 
    Both codes outperform the surface code by offering up to orders of magnitude lower logical error probabilities. 
    Interestingly, we find that this decrease in the logical error probability may not be accompanied by an increase in the code threshold, as different QEC codes benefit differently from large erasure fractions. 
    While here we focus on neutral atom qubits, the results also hold for other quantum platforms, such as trapped ions and superconducting qubits, for which erasure conversion has been demonstrated.
\end{abstract}

\maketitle

\section{Introduction}
Quantum error correction exploits redundancy to encode logical qubits in the state of many physical qubits, increasing robustness against physical errors when the error probability is below a given threshold \cite{gottesman2009introductionquantumerrorcorrection}. 
QEC invariably comes at the price of a large overhead in terms of resources, such as number of physical qubits, fidelity of operations, and control systems. This overhead is pushing the limits of experimental capabilities in all qubit architectures. The development of fault-tolerant strategies for QEC that can reduce the resource overhead is key to enabling practical and scalable quantum computation.

Surface \cite{Kitaev_2003,Dennis_2002,Fowler_2012} and color \cite{Bombin_2006,Bombin_2007} codes have so far been the leading paradigms to achieve QEC due to their high tolerance to errors, locality of stabilizer operators and two-dimensional layout, which can simplify experiments. Recently, using these codes, both logical Bell pairs preparation and  quantum memories with sub-threshold error probabilities have been demonstrated in neutral atom \cite{Bluvstein_2023} and superconducting \cite{Google2024} quantum processors, respectively. However, these codes suffer from a poor encoding rate (i.e., logical to physical qubit ratio) which poses severe limitations to scalability due to large qubit overhead.

Recently, there has been a surge of interest in designing hardware-tailored quantum Low-Density Parity-Check (LDPC) codes \cite{Bravyi2024,xu2023constantoverhead,pecorari2024highratequantumldpccodes,Berthusen:2024hit,poole2024architecturefastimplementationqldpc,PhysRevLett.129.050504} for QEC, namely stabilizer codes where the number of stabilizers acting on each qubit and the number of qubits acted on by each stabilizer are bounded by some constants \cite{gottesman2014faulttolerant,Breuckmann_2021}. These codes usually offer higher encoding rate and large code distance (the code parameter quantifying the maximum number of correctable errors) compared to the standard surface code, at the price of long-range connectivity. 

Bivariate Bicycle \cite{Bravyi2024} and La-cross \cite{pecorari2024highratequantumldpccodes} quantum LDPC codes are candidates for near-term experimental realizations in neutral atom \cite{pecorari2024highratequantumldpccodes,poole2024architecturefastimplementationqldpc,Berthusen:2024hit}, trapped-ion \cite{PhysRevLett.133.180601} and superconducting \cite{Bravyi2024,Berthusen:2024hit} platforms. They have been shown to offer significantly lower overhead and better error correction performance than the surface code \cite{Bravyi2024,pecorari2024highratequantumldpccodes}, while only requiring few non-local interactions of moderate extent.

Still, the implementation of high-rate quantum LDPC codes remains challenging, as these codes typically suffer from lower circuit-level thresholds than surface and color codes, as their higher stabilizer weight increases the number of error locations. 

QEC is also sensitive to the type of errors that occur. First, qubits engineered to be strongly biased towards a specific error mechanism have been demonstrated to allow for better suppression of errors at the logical level compared to unbiased qubits \cite{PhysRevA.78.052331,Lescanne_2020,Grimm2020StabilizationAO,PRXQuantum.2.030345,PhysRevX.12.021049}. Second, robustness against logical errors can be further enhanced by designing bias-tailored QEC codes that exploit noise biases \cite{Bonilla_Ataides_2021,Roffe_2023,Dua_2024}.
This bias-tailoring, also known as Clifford deformation \cite{Roffe_2023,Dua_2024}, consists of locally modifying the code stabilizers by applying single-qubit Clifford operators on the data qubits to create symmetries in the error patterns that dramatically simplify the decoding procedure. A well-known example is the XZZX surface code \cite{Bonilla_Ataides_2021}, i.e. a surface code where Hadamard rotations are performed on alternating data qubits, that performs particularly well against $Z$-biased noise. 

The XZZX surface code has also been shown to offer significant enhancements in QEC performance when a large fraction of errors are \emph{erasures} \cite{Wu_2022,Sahay_2023}, i.e. heralded qubit losses \cite{PhysRevLett.78.3217}. Erasures are less detrimental than generic depolarizing errors, since their actual location in the quantum register is known. The desirable scenario where qubits are strongly biased towards erasure errors can be experimentally realized via \emph{erasure conversion mechanisms} \cite{Wu_2022,PhysRevX.12.021049,PRXQuantum.5.040343,gu2024optimizingquantumerrorcorrection,PhysRevResearch.7.013249} to reveal the locations of losses. 
Examples of these mechanisms have been originally introduced to mitigate the effect of Rydberg leakages in Alkaline-earth(-like) atom qubits, such as Yb$^{171}$ \cite{Wu_2022,Sahay_2023,Ma_2023}, and atom losses \cite{PhysRevX.12.021049,PRXQuantum.5.040343,omanakuttan2024coherencepreservingleakagedetection,perrin2024quantumerrorcorrectionresilient}. Beyond neutral atom platforms, erasure conversion mechanisms have also been demonstrated for trapped ions \cite{Kang_2023,saha2024highfidelityremoteentanglementtrapped,quinn2024highfidelityentanglementmetastabletrappedion} and superconducting qubits \cite{PhysRevX.13.041022,Chou:2023kol,PhysRevX.14.011051}.

Identifying the \emph{best} QEC code family that can be implemented in near-term quantum processors is currently a key open challenge. Ideally, such a code family should have low overhead, high circuit-level threshold, and moderately long-range connectivity, while allowing for high logical error suppression. In this work, we move one step forward in this direction by showing that Clifford-deformed La-cross codes and, partially, Bivariate Bicycle codes fulfill all these requirements under a noise model strongly biased towards erasure errors. We consider two different types of erasure errors, namely \emph{unbiased erasures} and \emph{biased erasures} \cite{Wu_2022,Sahay_2023}. The former refers to the case where the erased qubits are reloaded in the the maximally mixed state, the latter to that where the erased qubits are reloaded in the first excited state, hence introducing -- upon Pauli twirling approximation -- an additional bias towards Pauli $Z$ errors. We use extensive QEC circuit-level numerical simulations with two-qubit gate errors and show that there exists an experimentally achievable range of physical error probabilities where both codes outperform the surface code in any respect, offering up to orders of magnitude lower logical error probabilities. In particular, Clifford-deformed La-cross codes offer thresholds that are comparable to those of the surface code under both unbiased and biased erasures. Instead, Bivariate Bicycle codes do not display significant threshold improvement for any finite fraction of unbiased erasures, while still offering competitive logical error probabilities below threshold. The noise model that we consider in our work can be experimentally realized in Alkaline-earth(-like) atom qubits \cite{Wu_2022}.

This work has two main implications. First, it demonstrates how new quantum LDPC code families, can significantly benefit from large erasure fractions, setting the stage for realizing low-overhead and high-threshold quantum LDPC code memories in near-term neutral atom quantum processors. Second, by comparing the performance of the La-cross and Bivariate Bicycle  code families, we prove that different codes can benefit differently from large erasure fractions, showing significant improvements either in both threshold and logical error probability or only in the logical error probability. 
These results open the way to the design of erasure-specific QEC codes that successfully trade between low logical error probabilities, high threshold, low overhead, and amenable long-range connectivity for near-term implementations. Although in this work we focus on neutral atoms, these results hold for other quantum platforms, such as trapped ions and superconducting qubits, for which similar erasure conversion mechanisms have also been demonstrated.

The remainder of this work is structured as follows. In Sec.~\ref{section:sec2} we review the paradigm of neutral atom quantum computing and provide an overview of the erasure conversion protocols in this platform. In Sec.~\ref{ssection:sec3A} we present Clifford-deformed La-cross codes 
and in Sec.~\ref{ssection:sec3B} Bivariate Bicycle codes. In Sec.~\ref{ssection:sec3C} we present error correction numerical simulations for both La-cross and Bivariate Bicycle code families, we discuss the main results for threshold and logical error probability improvements and compare them with the surface code. Afterwards, in Sec.~\ref{ssection:sec4A} we review the shuttling and static implementation schemes for quantum LDPC codes with neutral atom qubits. In Sec.~\ref{ssection:sec4B} we motivate in the experimental perspective the main assumptions of our erasure noise modeling, and finally in Sec.~\ref{ssection:sec4C} we outline some perspectives for the near-term realization of erasure conversion with quantum LDPC codes.

\section{Overview of erasure conversion protocols for neutral atoms}
\label{section:sec2}
In neutral atom quantum processors \cite{RevModPhys.80.885,RevModPhys.82.2313,Browaeys_2020,Henriet_2020,10.1116/5.0036562}, atoms -- generally Alkali or Alkaline-earth(-like) -- are loaded from a magneto-optical trap into a optical tweezer array generated by a spatial light modulator (SLM) and rearranged in a defect-free configuration via 2D acousto-optic deflectors (AODs) \cite{Bluvstein_2022}. Single-qubit rotations are performed via robust Raman laser-excitations with more than $99.95\%$ fidelity \cite{Bluvstein_2022}. Let atoms be three-level systems with computational states $\{|0\rangle,|1\rangle\}$ and auxiliary Rydberg state $|r\rangle$. Controlled-Z (CZ) entangling gates are usually realized by shining global laser pulses coupling the $|1\rangle\leftrightarrow|r\rangle$ states of each atom, such that the target atom in state $|1\rangle$ acquires a $\pi$-phase conditionally on the control atom also being in state $|1\rangle$ \cite{Jandura_2022}. The strong Van der Waals interaction between the two atoms naturally prevents them to be simultaneously Rydberg-excited \cite{RevModPhys.82.2313}. Recently, CZ gate fidelities in neutral atom qubits have recorded values of $99.4-99.8\%$ \cite{Evered_2023,Scholl_2023,peper2024spectroscopymodeling171ybrydberg,senoo2025highfidelityentanglementcoherentmultiqubit}. 

A particularly detrimental error source affecting neutral atom experiments is qubit loss. Several mechanisms can cause atoms to be lost during the computation, such as background gas collisions, heating by repeated gate execution or atom shuttling, and atom drift by imperfect trapping or imperfect recapture \cite{PhysRevX.12.021049,PRXQuantum.5.040343}. That is because, usually, optical traps are trapping when atoms are in their ground state and anti-trapping when they are in a Rydberg-excited state. Consequently, the optical traps have to be turned off during the gate execution and then turned on again, with a non-zero probability that the atom is lost during the process. Two other error sources causing qubit loss are Rydberg leakages and hyperfine leakages \cite{PRXQuantum.5.040343}. The former refers to atoms unintentionally left in some Rydberg state after the gate execution, either by over-rotation or by black-body induced transitions to neighboring Rydberg states. The latter refers to decays to hyperfine Zeeman sublevels. In both cases, atoms leave the computational subspace, resulting in information loss.

Most of these errors can be detected at the price of extra operations or extra ancillary atoms and converted into erasures. Atom losses and hyperfine leakages can be detected by fault-tolerantly integrating leakage detection unit circuits into the computation stack, as discussed in Ref.~\cite{PhysRevX.12.021049} for $^{87}$Rb and in Ref.~\cite{PRXQuantum.5.040343} for $^{137}$Cs atoms, and simulated for surface-code-based quantum error correction in Ref.~\cite{perrin2024quantumerrorcorrectionresilient}. Instead, Rydberg leakages can be converted into erasures via optical pumping \cite{PhysRevX.12.021049} or via mid-circuit ground state imaging of metastable Alkaline-earth(-like) atom qubits \cite{Wu_2022}.
 
In Alkaline-earth(-like) atoms, the computational subspace can be encoded into metastable long-lived clock states with lifetimes that can reach tens of seconds. The advantage of this encoding is that Rydberg leakages from gates mostly occur out of the computational subspace back to the ground state manifold of the system. Therefore, gate layers can be interleaved with mid-circuit ground state population measurements to reveal the location of the erased atoms without disturbing the qubit state, practically converting leakages into erasures. Erased atoms can then be re-pumped to the metastable states or reloaded from an external reservoir via movable tweezers. We note that this protocol makes possible to herald the exact failed gate along with the location of the erased atoms. Therefore, it both allows for extracting the maximum information possible about the errors and for correcting them \emph{on-the-fly} by reloading atoms during the computation. 
This protocol was first proposed in Ref.~\cite{Wu_2022} to mitigate Rydberg leakages in $^{171}$Yb qubits. Rydberg decay events during two-qubit gates mostly account for black-body-induced transitions to neighboring Rydberg states. Rydberg population after the gate can be removed via autoionization, ions can then be fluorescence-detected and finally removed by applying a small electric field. Raman-mediated qubit rotations similarly involve Rydberg-excited states and so single-qubit gate errors can analogously be converted into erasures. It has been theoretically demonstrated that up to $98\%$ of Rydberg decay events can be converted into erasures. Experimentally, $56\%$ of single-qubit gate errors and $33\%$ of CZ gate errors have been successfully converted into erasures to date \cite{Ma_2023}.

\begin{figure}[t]
    \centering
    \includegraphics[scale=1.04]{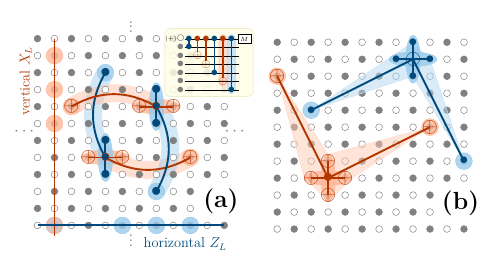}
    \caption{
    (a) Clifford-deformed La-cross quantum LDPC code ($k=3$ in this example) with two types of stabilizers mixing $X$ (red) and $Z$ (blue) Pauli operators. Logical operators are Pauli strings made of either all $X$ or all $Z$ operators (two examples are shown). The inset shows the syndrome extraction circuit for one stabilizer, which generalizes straightforwardly that of the XZZX surface code. (b) $[[72,12,6]]$ Bivariate Bicycle code, one $X$ (red) and one $Z$ (blue) stabilizer are shown.}
    \label{fig:layout}
\end{figure}

\section{Quantum LDPC codes}
In this section we review two recently introduced families of quantum LDPC codes, namely La-cross \cite{pecorari2024highratequantumldpccodes} and Bivariate Bicycle codes \cite{Bravyi2024}, that are good candidates for near-term implementations due to their moderate non-local connectivity requirements. We apply a Clifford deformation to La-cross codes that is analogous to that transforming the standard surface code into the XZZX surface code. We present and discuss error correction numerical simulations under both unbiased and biased erasure errors for La-cross codes and only unbiased erasure errors for Bivariate Bicycle codes.

\subsection{La-cross codes}
\label{ssection:sec3A}
We start by reviewing the hypergraph product construction and then generalize the Clifford deformation of the XZZX surface code \cite{Bonilla_Ataides_2021,Dua_2024,Roffe_2023} to arbitrary hypergraph product LDPC codes \cite{Tillich_2014,Kovalev_2013}, of which both La-cross and surface codes are examples. 

Let $\mathcal{C}_i=[n_i,k_i,d_i]$ be a classical code encoding $k_i$ logical bits in $n_i$ physical bits with Hamming distance $d_i$. Let $r_i$ be the number of its checks and $H_i\in\mathbb{F}_2^{r_i\times n_i}$ its parity-check matrix, i.e. the matrix having entries $(H_i)_{ab}=1$ iff the $a$th check acts non-trivially on the $b$th bit, so that $k_i=n_i-\text{rank}(H_i)$. Let $\mathcal{C}_i^T$ be the transposed code of $\mathcal{C}_i$ with parity-check matrix $H_i^T\in\mathbb{F}_2^{n_i\times r_i}$. We denote $\mathcal{C}_i^T\equiv[n_i^T,k_i^T,d_i^T]$ with obvious meaning of the code parameters. The hypergraph product (HGP) takes the parity-check matrices $H_i\in\mathbb{F}_2^{r_i\times n_i}$ of two classical codes $\mathcal{C}_i$ together with their transposed codes $\mathcal{C}_i^T$, $i=1,2$, and yields a $[[N,K,D]]$ quantum stabilizer code with quantum parity-check matrix
\begin{equation}
\label{eq:hgp_matrix}
H_Q=\left(\begin{array}{cc|cc}
        0 & 0 & H_1\otimes\mathbb{I}_{n_2} & \mathbb{I}_{r_1}\otimes H_2^T\\
        \mathbb{I}_{n_1}\otimes H_2 & H_1^T\otimes\mathbb{I}_{r_2} & 0 & 0
    \end{array}\right),
\end{equation}
where now $(H_Q)_{\alpha\beta}=1$ iff the $\alpha$th stabilizer acts non-trivially on the $\beta$th qubit. In Eq.~\ref{eq:hgp_matrix}, the left part of $H_Q$ describes $X$-stabilizers and the right one $Z$-stabilizers. Quantum parameters read $N=n_1n_2+r_1r_2$, $K=k_1k_2+k_1^Tk_2^T$, $D\geq\min\{d_1,d_2,d_1^T,d_2^T\}$ \cite{Tillich_2014} and the number of $X$- and $Z$-stabilizers is equal to $n_1r_2$ and $n_2r_1$, respectively. The resulting code is of Calderbank-Shor-Steane (CSS) type \cite{Calderbank_1996,Steane1996}, which means that the stabilizers of the code are products of only $X$ or only $Z$ Pauli operators. We can now perform a Clifford deformation consisting of applying Hadamard rotations to alternating data qubits \cite{Bonilla_Ataides_2021}. Upon this transformation, the above parity-check matrix becomes \cite{Roffe_2023}
\begin{equation}
\label{eq:deformed_matrix}
H_Q=\left(\begin{array}{cc|cc}
        0 & \mathbb{I}_{r_1}\otimes H_2^T & H_1\otimes\mathbb{I}_{n_2} & 0\\
        \mathbb{I}_{n_1}\otimes H_2 & 0 & 0 & H_1^T\otimes\mathbb{I}_{r_2}
    \end{array}\right).
\end{equation}
The resulting code has the same code parameters as the non-deformed one, but is no longer of CSS-type, as stabilizers now mix $X$ and $Z$ Pauli operators.

La-cross quantum LDPC codes are HGP codes built from equal cyclic classical codes with a generating polynomial $h(x)=1+x+x^k$, Hamming distance $d$, and classical parity-check matrix $H=\text{circ}(1,1,0,\dots,0,1,0,\dots,0)\in\mathbb{F}_2^{(n-k)\times n}$ (with the locations of the non-zero entries dictated by the generating polynomial, i.e. $0,1,k$) which is circulant rectangular \cite{pecorari2024highratequantumldpccodes}. It has been shown in Ref.~\cite{pecorari2024highratequantumldpccodes} that classical codes having circulant rectangular parity-check matrix guarantee the resulting La-cross codes to enjoy open boundary conditions and code parameters $N=(n-k)^2+n^2$, $K=k^2$, $D=d$. Additionally, La-cross codes have weight-$6$ stabilizers consisting of four nearest-neighbor and two symmetrically separated distant qubits -- or stabilizer \textit{legs}. The length of the stabilizer legs is equal to $k$. 
Since for La-cross codes the generating classical codes are identical, we can drop the indices and use the classical parameter $k$ as a label for different code instances. In fact, for equal rectangular seeds encoding $k$ bits, the resulting HGP code will encode $k^2$ logical qubits, hence the parameter $k$ sets both the long-range connectivity and the encoding rate of the resulting code. 

As in Ref.~\cite{pecorari2024highratequantumldpccodes}, in this work we focus on $k=2,3,4$ La-cross codes, corresponding to different degrees of moderately long-range connectivity. We show in Fig.~\ref{fig:layout}(a) a patch of a $k=3$ Clifford-deformed La-cross code. As opposed to the XZZX surface code, where the stabilizers are all equal, here we have two types of stabilizers, say \emph{left-} and \emph{right-handed}. However, given that $X$-legs are all horizontal and $Z$-legs are all vertical, these codes obey the same parity conservation law of the XZZX surface code under phenomenological noise, because all $Z$ ($X$) Pauli error strings are horizontally (vertically) aligned \cite{Bonilla_Ataides_2021}. For this reason, in the following we will not distinguish between left- and right-handed stabilizers.

\begin{figure*}[ht]
    \centering
    \includegraphics[scale=1.0]{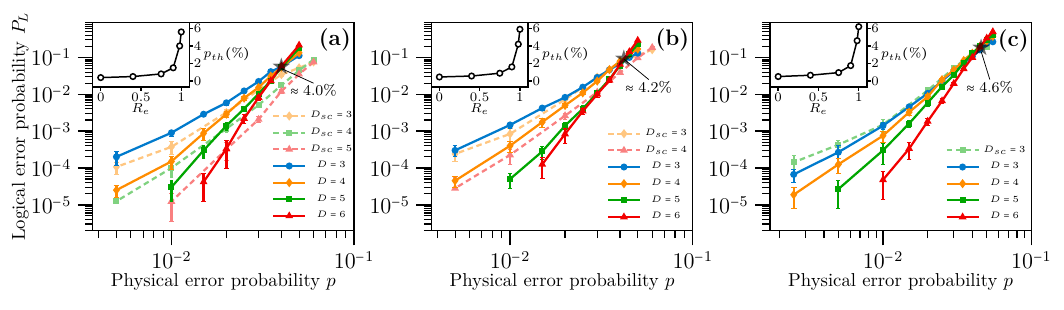}
    \caption{
    Cumulative logical error probability normalized by the number of QEC rounds for $k=2$(a), $k=3$(b) and $k=4$(c) La-cross codes (solid lines) under a fraction $R_e=0.98$ of unbiased erasure errors and $R_p=1-R_e=0.02$ of Pauli errors. A comparison with surface codes under the same noise model and at equal number of physical and logical qubits (dashed lines of the same color and marker style) is also shown. Codes sharing the same number of physical qubits, $N$, and logical qubits, $K$, are denoted with the same color. These results show that La-cross codes have high threshold (black stars) and significantly outperform the surface code in terms of QEC performance below physical error probabilities of $\sim10^{-2}$, for the code distances we have considered. The insets show how the thresholds of La-cross codes increase by increasing the unbiased erasure fraction for $R_e=0.40,0.75,0.90,0.98,1.0$. La-cross code instances shown in these plots are: $[[34,4,3]],[[52,4,4]],[[100,4,5]],[[130,4,6]]$ for $k=2$; $[[45,9,3]],[[65,9,4]],[[149,9,5]],[[225,9,6]]$ for $k=3$; $[[80,16,3]],[[106,16,4]],[[136,16,5]],[[208,16,6]]$ for $k=4$. Error bars on the data are standard deviations associated with the Monte Carlo error correction numerical simulations. For both La-cross and surface code, BP+OSD decoder was used.}
    \label{fig:code}
\end{figure*}

\subsection{Bivariate Bicycle codes}
\label{ssection:sec3B}
We now review the formal construction of Bivariate Bicycle codes. Let $I_\ell,S_\ell\in\mathbb{F}_2^{\ell\times\ell}$ be the identity matrix and the cyclic shift matrix, respectively, for some integer $\ell$. The quantum parity-check matrix of Bivariate Bicycle codes then reads \cite{Bravyi2024}
\begin{equation}
\label{eq:bb_matrix}
H_Q=\left(\begin{array}{cc|cc}
        0 & 0 & B^T &A^T\\
        A & B & 0 & 0
    \end{array}\right),
\end{equation}
where $A=A_1+A_2+A_3$ and $B=B_1+B_2+B_3$ are matrix trinomials with $A_i$ and $B_i$ powers of $x=S_\ell\otimes I_m$ and $y=I_m\otimes S_\ell$, respectively. Such a construction ensures the resulting code to be of CSS-type and to have periodic boundary conditions and weight-$6$ stabilizers, as $H_Q$ is square and each row contains only six non-zero entries. Bivariate Bicycle code parameters are: $N=2\ell m$, $K=2\times\text{dim}(\text{ker}(A)\cap\text{ker}(B))$, and $D=\text{min}\{|v|,\;v\in\text{ker}([A|B])\,\backslash\,\text{rowspace}([B^T|A^T])$ which can be computed using the integer linear programming method \cite{Bravyi2024}.

In this work, we focus on the following three code instances: $[[72,12,6]]$ with $(\ell,m)=(6,6)$ [see Fig.~\ref{fig:layout}(b)], $[[108,8,10]]$ with $(\ell,m)=(9,6)$ and $[[144,12,12]]$ (Gross code) with $(\ell,m)=(12,6)$, all three codes having $A=x^3+y+y^2$ and $B=y^3+x+x^2$. Additionally, in this case we do not perform any code deformation and benchmark the error correction performance offered by Bivariate Bicycle solely under unbiased erasure errors. That is because qubits in the support of the stabilizers (see Fig.~\ref{fig:layout}) are not all aligned along the same two directions of the array and therefore these codes do not enjoy the same parity symmetry discussed above for XZZX and La-cross codes.

\subsection{Error correction simulations}
\label{ssection:sec3C}
In this work we have evaluated the QEC performance of quantum LDPC codes via extensive numerical simulationsusing the Clifford simulator \texttt{Stim} \cite{gidney2021stim}. For Clifford-deformed La-cross codes, open boundary conditions, unrotated array configuration \cite{Fowler_2012,pecorari2024highratequantumldpccodes}, and syndrome extraction scheme depicted in Fig.~\ref{fig:layout}(a) (inset) are assumed \cite{pecorari2024highratequantumldpccodes}. Bivariate Bicycle codes are simulated with periodic boundary conditions, unrotated array configuration and standard syndrome extraction scheme. In both cases, the qubit register is first prepared to state $|0\rangle^{\otimes N}$ and -- only for Clifford-deformed La-cross codes -- Hadamard (H) rotations are applied on alternating data qubits. We apply CNOT/CZ gates between each data and ancilla qubit and then we measure the ancillas in the proper basis. To ensure fault tolerance against errors on the ancilla qubits \cite{Fowler_2012}, for all codes, the process is repeated $D$ times for a total of $D$ stabilizer measurement rounds, $D$ being the code distance (see Appendix A for a discussion about the optimal number of rounds.)

We focus on the effects of two-qubit gate errors, which are usually the most detrimental error mechanism, while we assume single-qubit and idling operations to be perfect. We also set the error strength of state preparation and measurement (SPAM) gates to zero, since we are only interested in evaluating two-qubit gate errors. Additionally, it was found in Ref.~\cite{Wu_2022} that including SPAM errors in the simulations only slightly decreases the surface code threshold, while the general sub-threshold behavior is left unchanged. Very similar considerations can be extended to the quantum LDPC codes considered here as well.

We also observe that the choice of neglecting single-qubit and idling errors is physically motivated in neutral atom qubits. In Ref.~\cite{Ma_2023}, the fidelity of single-qubit and idling operations are found to be mainly limited by the finite lifetime of the metastable states (few seconds). Specifically, single-qubit gate infidelities are typically more than one order of magnitude lower than two-qubit gates infidelities \cite{Evered_2023,Ma_2023}, while idle errors are negligible \cite{Bluvstein_2023,Ma_2023}. All these observations therefore motivate our choice of setting to zero the error strength of single-qubit and idling operations in our QEC numerical simulations.

We assume that each two-qubit gate can experience either a Pauli error with probability $p_p=p(1-R_e)$ or an erasure with probability $p_e=pR_e$, $R_e$ being  the fraction of erasure errors. Pauli errors are randomly drawn from $\{I,X,Y,Z\}^{\otimes2}\backslash\{I\otimes I\}$ with probability $p_p/15$, corresponding to the standard two-qubit depolarizing noise.  
Unbiased erasure errors are instead modeled as follows: When a two-qubit gate is erased, both atoms -- data and ancilla -- are reset to the maximally mixed state $I/2$, that is we draw errors at random from $\{I,X,Y,Z\}^{\otimes2}$ with probability $1/16$, conditionally on an erasure to have occurred. 
Experimentally, this is consistent with replacing the erased atoms with fresh ones initialized to the maximally mixed state from an external reservoir, e.g. via movable tweezers \cite{Wu_2022}. In addition, reloading both data and ancilla atoms involved in the erased gate prevents correlated errors to spread during the computation.

For infinitely $Z$-biased (\textit{biased} for short in this work) erasures, errors are drawn at random from $\{I,Z\}^{\otimes2}$ \cite{Sahay_2023}. This second noise model is physically motivated by the fact that in the standard Rydberg blockade gate, Rydberg excitations always occur when the qubit is in state $|1\rangle$ and never when it is in state $|0\rangle$. If no errors occur, after the gate the atom is de-excited back to state $|1\rangle$. Instead, if an erasure occurs, the atom leaks out of the computational subspace to some state $|e\rangle$. It can then be detected and replaced by a fresh one initialized to state $|1\rangle$. The combination of this quantum error channel with the described recovery operation results in a new quantum channel that is equivalent, via Pauli twirling approximation or randomized benchmarking, to the biased erasure quantum channel $\mathcal{E}(\rho)=1/2(I\rho I+Z\rho Z)$ \cite{Sahay_2023}.

For both noise models, we assume that mid-circuit ground-state measurements can be interleaved after any gate layer of the syndrome extraction circuit, ensuring that the failed CNOT/CZ gates are heralded along with the error locations and that the erased atoms can be reloaded, as we model in the simulations. We observe that, despite the increased time overhead of erasure conversion, the initial assumption of negligible idle errors is still well motivated. That is because experimentally errors are heralded via fast and destructive fluorescence imaging on the order of tens of microseconds, which is a much smaller time scale compared to that of idle errors, on the order of few seconds \cite{Ma_2023}.

\begin{figure*}[ht]
    \centering
    \includegraphics[scale=1.0]{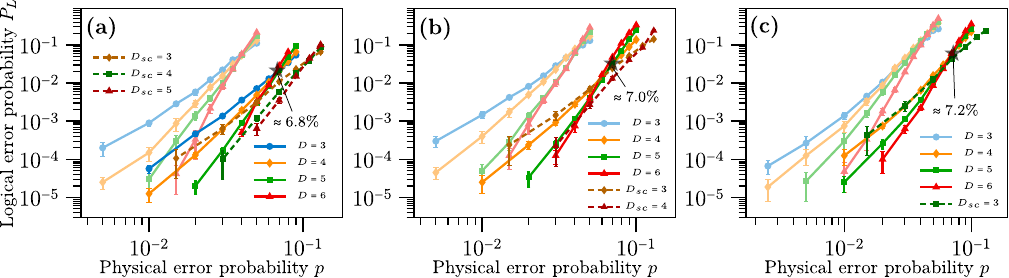}
    \caption{
    Cumulative logical error probability normalized by the number of QEC rounds for $k=2$(a), $k=3$(b) and $k=4$(c) La-cross code (colored lines) under a fraction $R_e=0.98$ of biased erasure errors and $R_p=1-R_e=0.02$ of Pauli errors. A comparison with La-cross codes under $R_e=0.98$ of unbiased erasure errors (pale lines) is shown, showing the benefit in the threshold (black stars) and in the logical error probability from a such a noise model. We also show a comparison with surface codes (dashed dark lines) with $R_e=0.98$ of biased erasure errors, at equal number of physical and logical qubits (dashed lines of the same color and marker style). Codes sharing the same number of physical qubits, $N$, and logical qubits, $K$, are denoted with the same color. La-cross codes start to outperform the surface code in terms of QEC performance below physical error probabilities of $\sim3\times10^{-2}$, for the code distances we have considered. La-cross code instances shown in these plots are: $[[34,4,3]],[[52,4,4]],[[100,4,5]],[[130,4,6]]$ for $k=2$; $[[45,9,3]],[[65,9,4]],[[149,9,5]],[[225,9,6]]$ for $k=3$; $[[80,16,3]],[[106,16,4]],[[136,16,5]],[[208,16,6]]$ for $k=4$. Error bars on the data are standard deviations associated with the Monte Carlo QEC simulations. For both La-cross and surface code, BP+OSD decoder was used.}
    \label{fig:bias_code}
\end{figure*}

\begin{table*}[hbt]
\centering
\begin{tabular}{cccc|ccc|ccc|ccc}
\toprule
\multicolumn{1}{c}{} & \multicolumn{3}{c}{Surface Code} & \multicolumn{3}{c}{La-cross $k=2$} & \multicolumn{3}{c}{La-cross \textbf{$k=3$}} & \multicolumn{3}{c}{La-cross \textbf{$k=4$}} \\
\cmidrule(rl){2-4} \cmidrule(rl){5-7} \cmidrule(rl){8-10} \cmidrule(rl){11-13} 
Model & {$R_e=0$} & {$R_e=0.98$} & {$R_e=1$} & {$R_e=0$} & {$R_e=0.98$} & {$R_e=1$} & {$R_e=0$} & {$R_e=0.98$} & {$R_e=1$} & {$R_e=0$} & {$R_e=0.98$} & {$R_e=1$} \\
\midrule
\makecell{circuit-level, \\ unbiased erasure} & \makecell{$1.6\%$ \\ ($1.1\%^*$)} & \makecell{$5.6\%$ \\ ($4.6\%^*$)} & \makecell{$6.8\%$ \\ ($5.4\%^*$)} & $0.40\%$ & $4.0\%$ & $5.6\%$ & $0.45\%$ & $4.2\%$ & $5.9\%$ & $0.50\%$ & $4.6\%$ & $6.2\%$\\
\cmidrule(rl){2-13}
\makecell{circuit-level, \\ biased erasure, \\ native gates} & \makecell{$1.6\%$ \\ ($1.1\%^*$)} & \makecell{$10.1\%$ \\ ($8.7\%^*$)} & \makecell{$11.7\%$ \\ ($10.6\%^*$)} & $0.40\%$ & $6.8\%$ & $9.1\%$ & $0.45\%$ & $7.0\%$ & $9.4\%$ & $0.50\%$ & $7.2\%$ & $9.6\%$ \\
\bottomrule
\end{tabular}
\caption{
Circuit-level thresholds normalized by the number of QEC rounds for XZZX surface code and Clifford-deformed $k=2,3,4$ quantum LDPC La-cross codes with unbiased and infinitely biased erasure errors in different fractions, namely $R_e=0.0,0.98,1.0$. For comparison with the literature, we also show surface code thresholds without normalization by the number of QEC rounds, which we mark with a ``*".}
\label{tab:thresholds}
\end{table*}

We perform extensive Monte Carlo numerical simulations to probe the code tolerance to errors and use minimum-sum Belief Propagation with Ordered Statistics Decoding (BP+OSD) \cite{Roffe_2020,Roffe_LDPC_Python_tools_2022} to process syndrome information of both quantum LDPC codes and surface codes. For all codes, we have set the number of iterations of Belief Propagation to $4-10$ and used Ordered Statistics Decoding at first order. These parameters have been found to optimally trade between decoder accuracy and time overhead \cite{pecorari2024highratequantumldpccodes}. Moreover, although significantly slower than standard matching-based decoders for the surface code, BP+OSD has proven to provide slightly higher threshold and lower logical failure rate \cite{pecorari2024highratequantumldpccodes}, therefore we have chosen to show here decoding plots for the surface code obtained with BP+OSD. We also mention that, while the protocol could significantly benfit form shorter decoding times, the huge time overhead of BP+OSD is not an immediate issue for erasure conversion in quantum memory experiments. In fact, error locations are heralded via fast ground state measurements and fresh atoms are reloaded in place of erased ones. Crucially, this does not require real-time decoding, as the erasure detection process is almost perfect, being only limited by the single atom imaging fidelity \cite{Ma_2023}. Therefore, for quantum memory experiments, offline decoding via BP+OSD represents a viable strategy. Instead, for deep quantum computation with, e.g., gate teleportation, real-time decoding is required and faster decoding strategies for quantum LDPC codes would then become necessary.

We normalize the data by the number of QEC rounds, $D$, i.e. we evaluate the logical failure probability as $P_L=1-(1-p_L)^{1/D}$, $p_L$ being the non-normalized cumulative error probability (that is, the probability that \emph{any} of the $K$ logical qubit fails), calculated as the ratio between the number of decoder failures and the total number of shots. For surface codes, the logical failure rate is computed as $P^K_L=1-(1-P_L)^{K}$, $K$ being the number of logical qubits of the compared quantum LDPC code \cite{pecorari2024highratequantumldpccodes}.

We show in Fig.~\ref{fig:code} results for $k=2$(a), $k=3$(b) and $k=4$(c) Clifford-deformed La-cross codes under a fraction of $R_e=0.98$ unbiased erasure errors. We plot their cumulative logical error probability, $P_L$, as a function of the physical error probability, $p$, for different code sizes (solid lines). We compare these codes with surface codes of equal/similar number of physical qubits and equal number of logical qubits under the same noise model (dashed lines with the same color and marker style).

We note that, since the encoding rate is the same for both La-cross and surface codes, $K/N_{LDPC}=K/N_{sc}^K$, with $N_{sc}^K=K\times N_{sc}$, the two codes share the same data and ancilla qubit overhead. In fact, the number of ancilla qubit is $N_{LDPC}^{anc}=N_{LDPC}-K$ for La-cross codes and $N_{sc}^{K,anc}=K\times(N_{sc}-1)=N_{sc}^K-K=N_{LDPC}^{anc}$ for surface codes.

It has been shown in Ref.~\cite{pecorari2024highratequantumldpccodes} that La-cross codes under depolarizing noise have circuit-level thresholds $p_{th}\approx0.4-0.5\%$, while for the surface code $p_{th}\approx1.6\%$. Here, our results show that, under large unbiased erasure fractions, the thresholds of the two code families are comparable. That is, we find thresholds of $p_{th}\approx4.0-4.6\%$ for La-cross codes [black stars in Fig.~\ref{fig:code}(a)-(c)] and $p_{th}\approx5.6\%$ for the surface code. This corresponds to an approximate improvement of $\times10$ against $\times5$, showing that the absolute threshold increase of La-cross codes is larger than that of the surface code. 

The comparison between La-cross and surface codes shows that there always exist physical error probabilities below which La-cross codes offer larger error suppression than the surface code [crossing points of solid and dashed lines of the same color in Fig.~\ref{fig:code}(a)-(c)]. In Ref.~\cite{pecorari2024highratequantumldpccodes}, La-cross codes display larger error suppression than the surface code at equal number of physical and logical qubits for physical error probabilities $p\lesssim10^{-3}$. Instead, in this work we find that, for a $R_e=0.98$ unbiased erasure fraction, the onset of improvement in the logical failure probability over the surface code is larger and occurs at higher and experimentally more accessible physical error probabilities. For example, for $D>5$ La-cross codes, the logical error probability at $p\approx10^{-2}$ is already approximately one order of magnitude lower than that of the surface code. This behavior is due to the higher thresholds and to the different dependence of the logical error probability on the code distance compared to pure depolarizing noise. In fact, for pure erasures, the scaling is $P_L\propto p^D$, which is a faster decrease of the logical error probability compared to depolarizing noise, for which $P_L\propto p^{\lfloor (D+1)/2\rfloor}$. For $R_e<1.0$, Pauli errors are dominant in the asymptotic limit, while erasures are dominant near-threshold. Therefore, we expect the logical error probability to scale faster around threshold, while asymptotically $P_L\propto p^{\lfloor (D+1)/2\rfloor}$ at $p\ll p_{th}$ \cite{Wu_2022} [for the bending visible in Fig.~\ref{fig:code}(a)-(c), see Appendix A].

Finally, analogously to the surface code \cite{Wu_2022}, we find that, by increasing the erasure fraction from $R_e=0.0$ to $R_e=1.0$, La-cross code thresholds first increase slowly and only get significantly high for very large erasure fractions, namely $R_e\gtrsim0.90$ [see Fig.~\ref{fig:code}(a)-(c), insets].

\begin{figure}[ht]
    \centering
    \includegraphics[scale=0.93]{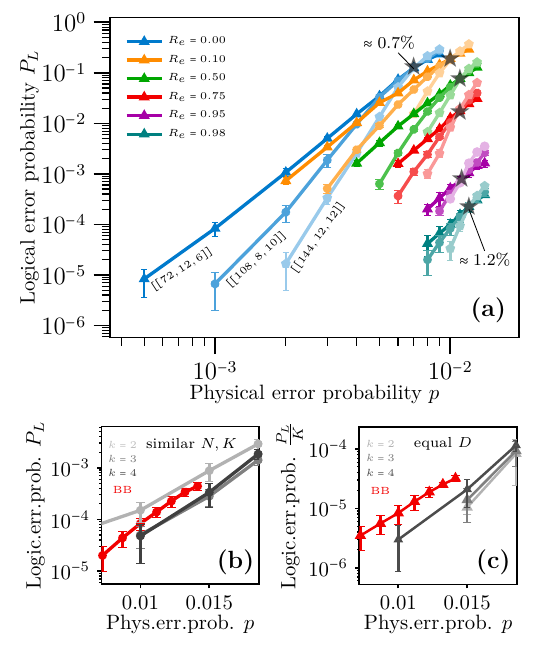}
    \caption{
    (a) Cumulative logical error probability normalized by the number of QEC rounds for Bivariate Bicycle (BB) codes under different fractions of unbiased erasure errors and Pauli errors, $R_e=0.00-0.98$. For each case, we study three code instances, namely $[[72,12,6]]$, $[[108,8,10]]$ and $[[144,12,12]]$, represented with different gradients of the same color and marker type. Thresholds (black stars) saturate at $\approx1.1-1.2\%$ for $R_e>0.10$, while the logical error probability at the threshold decreases. (b) Logical error probability at $R_e=0.98$ of unbiased erasures for Bivariate Bicycle (red) and La-cross codes (gray) at similar encoding rates, $R=K/N$. Codes considered are: $[[108,8,10]]$ ($R\approx0.074$) against $[[52,4,4]]$ ($R\approx0.077$), $[[149,9,5]]$ ($R\approx0.060$), and $[[208,16,6]]$ ($R\approx0.077$). In this case, BB codes are expected to quickly outperform La-cross codes below threshold due to the larger code distance at fixed qubit overhead ($D_{BB}=10$ against $D_{la-cross}=4,5,6$ in the shown example). (c) Logical error probability normalized by the number of logical qubits, $K$, at $R_e=0.98$ of unbiased erasures for BB (red) and La-cross codes (gray) at equal distance, $D=6$. La-cross codes slightly outperform BB codes due to the larger qubit overhead. In fact, La-cross encoding rates are lower than BB ones, i.e. $R_{BB}\approx0.167$, while $R_{k=2}\approx0.031$, $R_{k=3}=0.04$ and $R_{k=4}\approx0.077$ for $D=6$.}
    \label{fig:ibm_codes}
\end{figure}

The thresholds and the onset of improvement in the logical error probability for La-cross codes over the surface code get even larger under biased erasure errors. We show the decoding plots for $k=2,3,4$ La-cross codes with $R_e=0.98$ of biased erasures (colored lines) against unbiased erasures (pale lines), and surface codes $R_e=0.98$ of biased erasures at equal number of physical and logical qubits in Fig.~\ref{fig:bias_code}(a)-(c). The thresholds approximately increase from $4.0\%$ to $6.8\%$ for $k=2$ La-cross codes, from $4.2\%$ to $7.0\%$ for $k=3$ La-cross codes and from $4.6\%$ to $7.2\%$ for $k=4$ La-cross codes. These values are therefore again comparable to the $10.1\%$ threshold of the surface code under the same noise model, while the offered logical error protection is significantly larger. Consistently with the observations above for the unbiased erasure noise mode, also with biased erasures, La-cross codes outperform surface codes with equal number of physical and logical qubits for sufficiently small error probabilities, i.e. approximately below $p\lesssim10^{-2}$ for the code distances here considered.

To produce the plots with biased erasure errors, we have optimized the scaling factor of the Belief Propagation decoder routine, $s$, to minimize the output logical error probability, finding optimal values around $s=0.3-0.4$. We also note that, due to the lower accuracy of the decoder at large code sizes, we have chosen to identify as an upper bound for the thresholds the crossing point between the $D=4$ and $D=5$ lines [black stars in Fig.~\ref{fig:bias_code}(a)-(c)]. 

We observe that, despite the choice of considering Clifford-deformed La-cross codes, most of the gains from the unbiased and biased erasure noise models are found to be independent of the code deformation (we show the plots in Fig.~\ref{fig:fig_appendix}(b),(c) in Appendix A). In fact, for both CSS and non-CSS La-cross codes, we have not observed any difference in logical performance when enforcing the same noise model and the same decoding strategy via BP+OSD decoder. We note that this is consistent with our modeling of unbiased erasures as depolarizing errors with known location. On the other hand, although biased erasures are modeled as $Z$-biased errors, the use of native non-bias preserving gates effectively unbias the noise channel. We note that this behavior of La-cross codes is consistent with that of the surface code shown in Ref.~\cite{Sahay_2023}, where a comparable logical performance is observed for the XZZX and the standard surface code with $R_e=0.98$ of biased erasure errors.

We show in Tab.~\ref{tab:thresholds} a summary of the threshold values for surface and La-cross codes at $R_e=0.00,0.98,1.00$ for both unbiased and biased erasures. For the surface code, we also report in parentheses the threshold values that we have obtained before normalizing the data by the number of QEC rounds, for comparison with the literature (see Appendix B for a more detailed discussion). 

We find that, surprisingly, Bivariate Bicycle codes behave differently from La-cross codes under unbiased erasure errors, as shown in Fig.~\ref{fig:ibm_codes}(a), where the decoding plots for several erasure fractions are displayed. For $R_e=0.10$, the circuit-level threshold improves from $0.7\%$ ($R_e=0$) to $1.0\%$ and the logical error probability at the threshold, $P_L(p_{th})$, increases by a factor of approximately $\times1.5$. This behavior is similar to the one displayed by surface and La-cross codes with increasing unbiased erasure fractions. In contrast, for larger erasure fractions ($R_e=0.50,0.75,0.90,0.95$ in the figure) the threshold does not increase further, but rather seems to saturate around $p_{th}\approx1.1\%$ and the logical error probability at the threshold decreases. When $R_e=0.98$, the threshold is $p_{th}\approx1.2\%$, corresponding to an approximate $\times1.7$ improvement compared to pure depolarizing noise, and the logical error probability there is three orders of magnitude lower than that at $R_e=0.10$, namely $P_L(p_{th})\approx10^{-4}$.

We now compare Bivariate Bicycle codes against La-cross codes in terms of logical error correction performance in the sub-threshold regime when most errors are erasures, namely $R_e=0.98$. In Fig.~\ref{fig:ibm_codes}(b), we compare one exemplary instance of Bivariate Bicycle codes, namely the $[[108,8,10]]$ code, against La-cross codes of similar encoding rate, $R=K/N$, namely the $[[52,4,4]]$, $[[149,9,5]]$ and $[[208,16,6]]$ codes. We find that below threshold, the Bivariate Bicycle code quickly outperforms all the La-cross codes considered for $p\lesssim0.008$, due to the larger code distance responsible for a steeper logical error scaling. 
In Fig.~\ref{fig:ibm_codes}(c), we instead compare the logical error protection offered by the $D=6$ Bivariate Bicycle code, i.e. $[[72,12,6]]$ at $R_e=0.98$ of unbiased erasures with that offered by $k=2,3,4$ equal distance ($D=6$) La-cross codes, i.e. $[[130,4,6]],\,[[225,9,6]],\,[[2018,16,5]]$, under the same noise model. The logical error probability of Bivariate Bicycle codes normalized by the number of logical qubits (red line) is slightly higher than that of all La-cross codes (gray lines). That is because below threshold the QEC performance improves with the system size and La-cross codes have a larger number of physical qubits at the same distance. In fact, the encoding rate, $K/N$, offered by Bivariate Bicycle codes is larger, 
although still asymptotically vanishing, $K/N\xrightarrow{N\rightarrow\infty}0$, hence the overhead is lower. The error correction performance of these codes at equal distance is therefore comparable in the range of physical error probabilities $p\lesssim1\%$. 

In conclusion, Bivariate Bicycle codes offer a lower threshold than La-cross codes when most errors are erasures. At the same time, erasure conversion comes at the cost of large operation and time overheads, which make higher threshold desirable to enable near-term implementations. This suggests that in near-term noisy quantum devices La-cross codes might be a more promising candidate to realize erasure conversion in quantum LDPC codes. However, the comparison in logical error correction performance shows that Bivariate Bicycle codes allow to target comparable or lower logical error probabilities with significantly lower qubit overhead. Therefore, these codes might ultimately be preferred at sufficiently small physical error probabilities, approximately $p\lesssim8\times10^{-3}$. Additionally, we observe that the value of $P_L(p_{th})\approx10^{-4}$ offered by Bivariate Bicycle codes under $R_e=0.98$ of unbiased erasures is far below break-even (i.e., $P_L(p)=p$) and therefore potentially interesting  also for above-threshold near-term QEC experiments.

\begin{figure}[t]
    \centering
    \includegraphics[scale=.75]{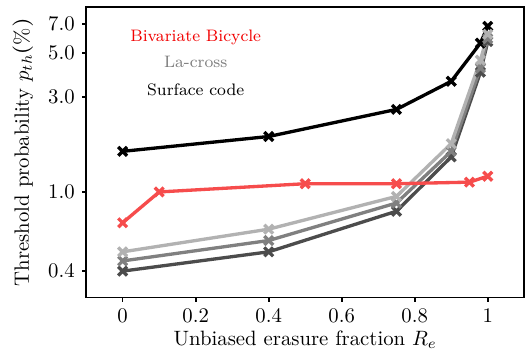}
    \caption{Threshold probabilities for different fractions, $R_e$, of unbiased erasure errors of Bivariate Bicycle (red), $k=2,3,4$ La-cross (gray) and surface codes (black). For La-cross codes, despite the low values for depolarizing errors, the threshold rapidly increase with the increasing erasure fraction and for $R_e\gtrsim0.98$ they become comparable to the high surface code thresholds. Instead, for Bivariate Bicycle codes, the threshold saturates at $p_{th}^{BB}\approx1.1\%$ and does not display any significant improvement for increasing erasure fractions.}
    \label{fig:fig5}
\end{figure}

Finally, our results show that different quantum LDPC code families can benefit differently from large erasure fractions. In particular, Bivariate Bicycle codes still do take advantage of large fractions of erasure errors, like surface and La-cross codes, but this advantage manifests itself only as a lower logical error probability, without any significant threshold improvement, as summarized in Fig.~\ref{fig:fig5}. 
The reason for not observing any significant increase in the threshold for Bivariate Bicycle codes is likely due to the fact that, under pure depolarizing noise, these codes already show high values of $D\times P_L(p_{th})$ (the $D\times$ factor standing for before round normalization), which gets $D\times P_L(p_{th})\approx1$ at $R_e=0.10$ (cf. Fig.~\ref{fig:ibm_codes}(a)) and therefore cannot increase further. However, a noise model with a large erasure fraction has a lower noise entropy and is easier to decode than pure depolarizing noise. Therefore, the codes are still expected to offer larger error suppression, i.e. lower logical error probability, as we observe in this work. 

Decoding plots of Bivariate Bicycle codes have been obtained by optimizing over the Belief Propagation scaling factor to find the lowest logical error probabilities, finding optimal values for $s=0.3-0.4$.

\section{Near-term implementation of quantum LDPC codes}
\label{section:sec4}
In this section, we review two proposed implementation schemes for quantum LDPC codes, namely the dynamic scheme via atom shuttling and the static scheme via long-range interaction engineering. We use that to motivate the assumptions of equal erasure fraction for all gates and range-independent error probability made in the previous section. In fact, both schemes are promising for implementing quantum LDPC codes in the near term, and both can be enhanced with erasure conversion for a better error correction performance. 

\subsection{Implementation schemes for quantum LDPC codes with neutral atoms}
\label{ssection:sec4A}
Implementing high-rate quantum LDPC codes in current noisy intermediate-scale quantum (NISQ) hardware is challenging. In fact, all quantum LDPC codes suffer from a lower degree of parallelism and a generally higher control complexity at the hardware level, although offering lower overhead and logical error probabilities compared to the surface code. That is mainly due to the large number of gates making up their stabilizers, some of them being long-range, which requires extra connectivity-enhancing resources. 
There are two leading paradigms for the short-term implementation of quantum LDPC codes in two-dimensional neutral atom arrays: The \emph{dynamic} scheme via qubit shuttling \cite{xu2023constantoverhead}, and the \emph{static} scheme via long-range interaction engineering \cite{pecorari2024highratequantumldpccodes,poole2024architecturefastimplementationqldpc}. Although we do not discuss it in this work, we mention that another strategy is to exploit cavity-mediated long-range interactions \cite{PhysRevA.110.062610,chandra2024nonlocalresourceserrorcorrection,srivastava2025cavitypolaritonblockadenonlocal,grinkemeyer2024errordetectedquantumoperationsneutral}.

In the implementation via qubit shuttling, atoms are trapped in movable tweezer traps generated by a crossed 2D acousto-optic deflector and then translated in parallel to execute CZ gates between neighboring data-ancilla atom pairs \cite{xu2023constantoverhead}. This allows for any-to-any connectivity and nearest-neighbor gate execution at the price of large numbers of rearrangement steps and long QEC cycle times, on the order of tens of milliseconds, mostly dominated by atom rearrangement time \cite{xu2023constantoverhead}. 

The static implementation paradigm requires engineering long-range interactions using multiple lasers to enable CZ gates between distant qubits. Transitions to Rydberg-excited energy levels with different higher principal quantum numbers are required to execute gates over different interatomic distances \cite{pecorari2024highratequantumldpccodes}. In fact, using higher principal quantum number Rydberg states enables larger blockade radii that are necessary to implement long-range gates. At the same time, larger spatial blockades result in larger Rydberg cross-talk and lower parallelism.
Even when accounting for this lower parallelism, this scheme is estimated to be as much as one order of magnitude faster than atom shuttling, although the CZ gate fidelity decreases quasi-linearly with the gate range \cite{pecorari2024highratequantumldpccodes}. The protocol can be further sped up by loading atoms in a folded array configuration to effectively reduce the gate extents. As a consequence, gate infidelities and durations decrease. Array folding has been originally proposed in Ref.~\cite{poole2024architecturefastimplementationqldpc} for Bivariate Bicycle codes, and it also applies to La-cross codes with open boundary conditions. La-cross codes with periodic boundaries and the standard Shor syndrome extraction scheme would instead suffer from hook errors, as demonstrated in Ref.~\cite{pecorari2024highratequantumldpccodes}.

\subsection{Equal erasure fraction and range-independent error strength}
\label{ssection:sec4B}
In this work we have mostly focused on erasure conversion to mitigate Rydberg decays in Alkaline-earth(-like) atom quantum processors. That is because single- and two-qubit gate fidelities in neutral atom qubits are mostly decay-limited, to date. We now motivate two assumptions we have made in the previous section, namely equal erasure fraction, $R_e$, for all stabilizer gates and range-independent two-qubit gate error probabilities, $p_e$ and $p_p$.

The assumption of equal erasure fraction $R_e$ for all gates can be motivated for both the dynamic and static realization schemes.
In the implementation via qubit shuttling, all CZ gates are executed nearest-neighbor. Therefore, if qubit shuttling is supplied with erasure conversion, all gate errors can be modeled as in Ref.~\cite{Wu_2022} and up to $98\%$ of them can be converted into erasures. Instead, in the static implementation, it is still in principle possible to remove all the Rydberg population after the gate, either by waiting a longer time (due to the longer Rydberg lifetime of higher-excited states) or via autoionization as discussed in Ref.~\cite{Wu_2022} for short-range gates, once again converting up to $98\%$ of gate errors into erasures.

The assumption of range-independent erasure probability, $p_e$, holds true for the shuttling scheme, where all gates are nearest-neighbor. On the other hand, for the static implementation scheme, the range-independent assumption is approximately consistent with quantum LDPC codes having moderate long-range connectivity. That is because long-range gates involve Rydberg states with only slightly larger principal quantum numbers, $n$, than those of nearest-neighbor gates ($n\approx50-60$ for nearest neighbor to next-to-nearest neighbor connectivity \cite{pecorari2024highratequantumldpccodes}).  
The range-dependent assumption is instead physically motivated for non-erasure-like errors from long-range gates \cite{pecorari2024highratequantumldpccodes}, although in this work we have modeled these errors with range-independent depolarizing noise. That is because it has been shown that, for moderate long-range connectivity, such as that of La-cross and Bivariate Bicycle codes, the range-dependence assumption does not significantly affect the code threshold and the error correction performance \cite{pecorari2024highratequantumldpccodes}. Moreover, in most of the cases discussed in this work, these errors are expected to play little role as they come in small fractions, i.e. $R_p=1-R_e=0.02$.

\subsection{Erasure conversion with quantum LDPC codes}
\label{ssection:sec4C}
We here discuss the possibility of equipping quantum LDPC codes with erasure conversion for Rydberg decays.

The theoretical erasure conversion scheme requires interleaving ground state measurements after any gate layer of the syndrome extraction circuit. This guarantees that the specific gate that failed is heralded with the erased qubit location, thus maximizing the noise information and dramatically simplifying the decoding problem. However, integrating erasure conversion into the compilation of a full QEC code -- where large numbers of gates have to be executed to correct for errors -- is hard. This is due to the fact that quantum LDPC codes typically suffer from a lower degree of parallelism compared to, for example, the surface code.
That is detrimental for erasure conversion, as it would require, e.g. massive ground state fluorescence imaging and atom shelving. As a result, the total time budget and physical error probability would increase. 

The problem of parallel compilation of quantum LDPC codes can be partially mitigated by optimizing the parallelism of the compiling circuits in a hardware-aware manner.
For example, the parallelism of the shuttling implementation scheme is established by the minimum number of trap movements required to implement the stabilizer measurement circuit and is therefore geometrically restricted. In contrast, in the static case, it is the larger Rydberg blockade radius due to excitations to high principal quantum number states that prevents the parallel execution of gates \cite{pecorari2024highratequantumldpccodes}. This issue could be improved by adopting a different gate scheme, e.g. the dark-state-mediated gate described in Ref.~\cite{Petrosyan_2017}, or by exploiting the anisotropy of the Van der Waals interaction \cite{Barredo_2014,wadenpfuhl2024unravellingstructuresvander}. We defer the problem of the optimal compilation of QEC codes in neutral atom registers to future work. 

Another possibility is to perform less frequent ground state imaging, for example, after two gate layers, or after each QEC round. In this case, measuring an atom in its ground state would solely herald the location of the leaked qubit and not the failed CZ gate. As a consequence, less information is transferred to the decoder, leading to lower logical error correction performance. This scenario is similar to that of atom loss errors \cite{PhysRevX.12.021049,PRXQuantum.5.040343,perrin2024quantumerrorcorrectionresilient}, with the mid-circuit ground state measurement now playing the role of a leakage detection unit (LDU). The expected scaling of the logical error probability for uncorrelated data-ancilla loss has been shown to be the same as the one of erasure errors, that is $P_L(p)\propto p^D$, although the threshold for the surface code has been proven to be lower ($2.6\%$ for pure losses) \cite{perrin2024quantumerrorcorrectionresilient}.

\section{Conclusions}

In this work, we have discussed two families of quantum LDPC codes, namely Clifford-deformed La-cross codes and Bivariate Bicycle codes, and performed extensive error correction numerical simulations under mixtures of Pauli and unbiased/biased erasure errors. We have mostly focused on the case $R_e=0.98$ for applications to Alkaline-earth(-like) atom qubits. With respect to pure depolarizing noise, for La-cross codes our results show a large increase in the circuit-level thresholds, which are comparable to those offered by the surface code under the same noise models. We also show significant improvements in the error correction performance of La-cross codes over the surface code, with more than one order of magnitude gain for experimentally achievable physical error probabilities, $p\lesssim10^{-2}$. 
On the other hand, we show that Bivariate Bicycle codes do not offer any significant improvement in the threshold, which is $p_{th}\approx1\%$ for any fraction of unbiased erasure errors. However, these codes still display a large improvement in the logical error probability, which is comparable to that of equal-distance La-cross codes under the same noise model. 

These results show how the error correction performance of new quantum LDPC code families can benefit from large erasure fractions, as originally demonstrated for the surface code, setting the stage for near-term experimental demonstrations of high-threshold and low-overhead quantum LDPC code memories.  
Moreover, these results prove that different code families can benefit differently from large erasure fractions and that the main feature of erasure conversion is the drastic diminishment of the logical error probability, also due to erasures being more amenable to decode than Pauli errors, rather than an increase of the circuit-level threshold, as mostly believed. This pushes the need for identifying erasure-specific QEC resources, not only at the physical level (e.g., fast or robust quantum gates \cite{Jandura_2022,PRXQuantum.4.020336,jandura2024surfacecodestabilizermeasurements}), but also at the logical level, by designing QEC codes that offer optimal performance under erasure errors, trading between overhead reduction, high threshold, good error correction capabilities, and amenable long-range connectivity for near-term implementations.

Although in this work we have mostly focused on erasure conversion to mitigate Rydberg decays in Alkaline-earth(-like) atom qubits, these results may be generalized to other quantum platforms for which erasure conversion protocols have also been demonstrated, such as trapped ions \cite{Kang_2023,saha2024highfidelityremoteentanglementtrapped,quinn2024highfidelityentanglementmetastabletrappedion} and superconducting qubits \cite{PhysRevX.13.041022,Chou:2023kol,PhysRevX.14.011051}, and, partially, to other error mechanisms that can be similarly converted into erasures, such as atom loss \cite{PhysRevX.12.021049,PRXQuantum.5.040343,omanakuttan2024coherencepreservingleakagedetection,perrin2024quantumerrorcorrectionresilient} or mixtures of atom loss and Rydberg decay \cite{postema2025geometricalapproachlogicalqubit}.

\begin{figure*}[ht]
    \centering
    \includegraphics[scale=1.0]{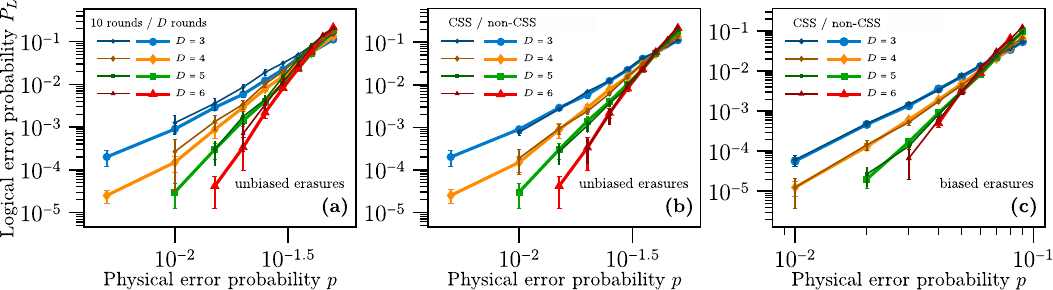}
    \caption{ 
    Supplementary details on QEC numerical simulations. Cumulative logical error probability normalized by the number of QEC rounds for $k=2$ La-cross codes. (a) Decoding plots performing either $D$ (colored thick lines) or $10$ rounds (dark thin lines) of stabilizer measurements. Except for the smallest code shown, with $D=3$, the values of the logical error probabilities are comparable within the error bars. (b) QEC simulations with $R_e=0.98$ of unbiased erasure errors for Clifford-deformed non-CSS La-cross codes (colored thick lines) and the standard CSS La-cross codes (dark thin lines) with $D$ rounds of stabilizer measurements. Simulations show no difference in performance for the two codes, consistently with the fact that the noise model is unbiased. (c) QEC simulations with $R_e=0.98$ of biased erasure errors for Clifford-deformed non-CSS La-cross codes (colored thick lines) and the standard CSS La-cross codes (dark thin lines) with $D$ rounds of stabilizer measurements. Simulations show no difference in performance for the two codes, suggesting that at circuit-level, non-bias preserving gates completely unbias the error channel. Code instances shown in these plots are: $[[34,4,3]],[[52,4,4]],[[100,4,5]],[[130,4,6]]$. Error bars on the data are standard deviations associated with the Monte Carlo QEC simulations. BP+OSD decoder was used.}
    \label{fig:fig_appendix}
\end{figure*}

\section*{Appendices}
\subsection*{Appendix A: Details on QEC simulations}
In this section, we provide additional details to support the QEC numerical simulations shown in the main text.

\subsubsection{Stabilizer measurement rounds}
In the main text, we have shown simulations for surface, La-cross and Bivariate Bicycle codes by performing $D$ rounds -- as many as the code distance -- of stabilizer measurements. This guarantees robustness against measurement errors with good approximation \cite{Fowler_2012}. However, the logical error probability is sensitive to the number of rounds \cite{PhysRevLett.129.050504}. For a specific code instance, namely Clifford-deformed $k=2$-La-cross codes with $98\%$ of unbiased erasure errors, we show in Fig.~\ref{fig:fig_appendix}(a) QEC simulations with $10$ (dark thin lines) and $D$ rounds (colored thick lines). Except for the shorter distance code shown ($D=3$), the logical error probabilities are consistent within the error bars, while crucially the threshold remains the same. This shows that the threshold values presented in the main text are independent of the number of rounds used in the simulations.

\subsubsection{Noise bias and code bias-deformation}
We now discuss the relation between erasure noise models and bias deformation in more detail. To do so, for the standard CSS non-deformed La-cross codes we repeat the same QEC simulations that we have shown in the main text for their Clifford-deformed counterparts. We focus only on the case $k=2$, with a fraction $R_e=0.98$ of both unbiased and biased erasure errors. For $k=3,4$ La-cross codes a similar behavior is expected. We show in Fig.~\ref{fig:fig_appendix}(b),(c) the resulting plots for unbiased and biased erasures, respectively. In both cases, the simulations with BP+OSD decoder show no difference in logical performance between CSS and non-CSS codes with both noise models. Unbiased erasures are simulated and decoded as unbiased depolarizing errors with known locations, therefore we expect no difference between CSS and non-CSS La-cross codes, as we correctly observe. Biased erasure are instead simulated and decoded as $Z$-biased errors with known locations, and bias-deformed codes are known to perform better under biased code-capacity and phenomenological noise \cite{Bonilla_Ataides_2021}. However, in the present case, simulations are performed at circuit level with native non-bias-preserving gates. The comparable performance of CSS and non-CSS La-cross codes then signals that the noise channel gets effectively unbiased during the computation when native gates are used. We observe that this behavior is not surprising, as it has previously been observed for the XZZX and standard surface codes in Ref.~\cite{Sahay_2023}.

Since all the simulations shown in this work have been performed with the same decoding strategy for both CSS and non-CSS codes, the above observations incidentally also validate the comparison between non-CSS La-cross codes and CSS Bivariate Bicycle codes with unbiased erasure errors and the same decoder presented in the main text.

\subsubsection{Surface code decoding strategies}
Finally, we comment on our choice of decoding surface codes with BP+OSD decoder, instead of using faster matching-based decoders. Upon comparing the performances of BP+OSD and \texttt{pymatching} \cite{Higgott2025sparseblossom} decoders on the surface code, BP+OSD was found to yield comparable logical error probabilities within the error bars and slightly higher thresholds than \texttt{pymatching}, and so the former was preferred in this context. Although the slightly higher accuracy of BP+OSD comes at the price of a larger time overhead, in this work we are interested in assessing the QEC performance of different quantum memories, which can be efficiently decoded \emph{offline}. In fact, erasure conversion does not require \emph{real-time} decoding. That is because the fidelity of the measurements that herald the errors is solely limited by the single-atom fluorescence imaging fidelity, which is practically $1$, with good approximation. In Ref.~\cite{Ma_2023}, no scattering errors have been observed in the erasure detection process, thus bounding their probability to $<10^{-6}$ per imaging time.

We conclude by discussing the bending shown at low physical error probabilities by the decoding curves of small distance codes, e.g., $D=3,4$, (Fig.~\ref{fig:code},~\ref{fig:bias_code}). 

\begin{figure}[ht]
    \centering
    \includegraphics[scale=0.9]{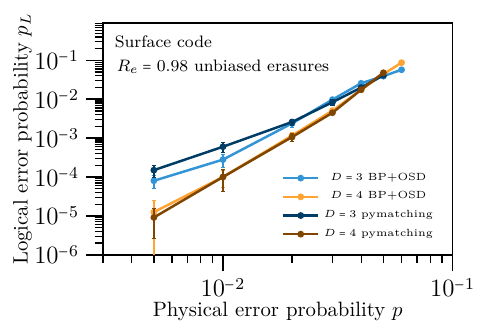}
    \caption{ 
    Surface code QEC simulations with BP+OSD and \texttt{pymatching} decoders when $98\%$ of the errors are unbiased erasures. Overall, the two decoders performs comparably within the errors bars, while BP+OSD yields a slightly higher error threshold and lower logical error probability for $D=3$. This shows that BP+OSD slightly outperforms, or performs comparably to, \texttt{pymatching} in the simulated regime of physical error probabilities. The logical error rate is here not normalized by the number of QEC rounds. Error bars on the data are standard deviations associated with the Monte Carlo QEC simulations.}
    \label{fig:fig_appendix2}
\end{figure}

We have performed numerical simulations for the surface code in order to compare the decoding performance of BP+OSD and \texttt{pymatching} decoders.  
We show in Fig.~\ref{fig:fig_appendix2} the result of the decoder comparison for the $[[13,1,3]]$ and $[[25,1,4]]$ surface codes. Although BP+OSD slightly overestimates the logical error probability near-threshold (which could be fixed by adjusting the BP scaling factor in that regime), \texttt{pymatching} shows no bending, but performs comparably or slightly worse within the error bars in the deep sub-threshold regime. In addition, the threshold probability obtained by \texttt{pymatching} is slightly lower than that obtained by BP+OSD (see Appendix B below). We conclude that BP+OSD is a valuable decoding strategy for the surface code in the simulated regime of physical error probabilities.

\subsection*{Appendix B: Surface code thresholds}
We note that in this work we have found surface code thresholds that are slightly larger than those known from the literature. That is marginally due to the different decoder we have used, namely BP+OSD, but mostly to the normalization by the number of QEC rounds. For consistency, here we report the values that we have obtained before round normalization together with the literature values. For unbiased erasures, we find surface code thresholds of $4.6\%$ for $R_e=0.98$ (Cfr. $4.15\%$ with Union Find decoder in Ref.~\cite{Wu_2022} and $4.3\%$ with Minimum Weight Perfect Matching decoder in Ref.~\cite{Sahay_2023}) and $5.4\%$ for $R_e=1.0$ (Cfr. $5.13\%$ with Union Find decoder in Ref.~\cite{Wu_2022} and $5.0\%$ with Minimum Weight Perfect Matching decoder in Ref.~\cite{Sahay_2023}). Instead, for infinitely biased erasures, we find surface code thresholds of $8.7\%$ for $R_e=0.98$ (Cfr. $8.2\%$ with Minimum Weight Perfect Matching decoder in Ref.~\cite{Sahay_2023}) and
$10.6\%$ for $R_e = 1.0$ (Cfr. $10.3\%$ with Minimum Weight Perfect Matching decoder in Ref.~\cite{Sahay_2023}). We conclude that these results are consistent up to decoder precision.

\section*{Acknowledgements}
We thank Hugo Perrin for many fruitful discussions and comments on the manuscript. We also gratefully acknowledge discussions with Gavin K. Brennen and Sven Jandura. This research has received funding from the European Union’s Horizon 2020 research and innovation programme under the Horizon Europe programme HORIZON-CL4-2021-DIGITAL-EMERGING-01-30 via the project 101070144 (EuRyQa) and from the French National Research Agency under the Investments of the Future Program projects ANR-21-ESRE-0032 (aQCess), ANR-22-CE47-0013-02 (CLIMAQS), and ANR-22-CMAS-0001 France 2030 (QuanTEdu-France).  
Computing time was provided by the High-Performance Computing Center of the University of Strasbourg. Part of the computing resources were funded by the Equipex Equip@Meso project (Programme Investissements d'Avenir) and the CPER Alsacalcul/Big Data.

\bibliography{bibliography}

\end{document}